\documentclass[aps,prl,twocolumn,superscriptaddress,10pt]{revtex4-2}

\usepackage[english]{babel}
\usepackage{amsmath}
\usepackage{float}  

\usepackage{physics}   
\usepackage{braket}    
\usepackage{graphicx}
\usepackage{eucal}
\usepackage[colorlinks=true, allcolors=blue]{hyperref}
\usepackage{ulem}
\usepackage{xcolor}
\usepackage{ulem}

\usepackage{xcolor}
\usepackage[markup=underlined]{changes}

\newcommand{\eq}[1]{Eq.~(\ref{eq:#1})}

\begin{document}

\title{Mediated Interactions between Fermi Polarons in the Strongly Interacting Regime}

\author{Renato Pessoa}
\affiliation{Instituto de F\'{\i}sica, Universidade Federal de Goi\'as - UFG, 74001-970 Goi\^ania, GO, Brazil}

\author{S.~A.~Vitiello}
\affiliation{Instituto de F\'{\i}sica Gleb Wataghin, Universidade Estadual de Campinas - UNICAMP, 13083-970 Campinas, SP, Brazil}

\author{L.~A.~Pe\~na Ardila}
\email{luis.ardila@units.it}

\affiliation{
Dipartimento di Fisica, Universit\`a di Trieste, 
Strada Costiera 11, I-34151 Trieste, Italy
}

\affiliation{
Istituto Nazionale di Fisica Nucleare (INFN), 
Trieste Section, Via Valerio 2, I-34127 Trieste, Italy
}

\begin{abstract}
Mediated interactions between quasiparticles lie at the heart of Landau Fermi-liquid theory, yet their behavior in the strongly correlated regime is not fully understood. Motivated by recent experiments [Nat. Phys. 20, 68–73 (2024)], we investigate mediated interactions between quasiparticles emerging from a Fermi sea—Fermi polarons using zero temperature quantum Monte Carlo methods. Our main result is that the Landau interaction between polarons remains repulsive in the strongly interacting regime for any coupling strength in the polaronic branch, in contrast to the downward trend observed experimentally near resonance. Deep in the short-range universal regime, our results are insensitive to the microscopic details of the interaction potential, providing robust evidence for the sign of the ground-state induced interaction. In addition, we determine the real-space interaction between static impurities, recovering a Ruderman–Kittel–Kasuya–Yosida (RKKY)-like behavior at weak coupling and the nonperturbative Born–Oppenheimer result at strong coupling. More broadly, our results establish a microscopic benchmark for how Landau quasiparticle interactions survive beyond the perturbative regime of strongly correlated quantum matter.

\end{abstract}

\maketitle

\textit{Introduction-- }Mediated interactions are a  key mechanism behind emergent many-body phenomena~\cite{Anderson1972MoreIsDifferent}. They arise through exchanged or collective excitations, such as gauge bosons in fundamental interactions, phonons in conventional superconductivity~\cite{Cooper1956,BCS1957}, and spin waves in magnetic systems~\cite{Berk1966}. At low energies, typically much smaller than the Fermi energy, these microscopic mediating mechanisms manifest effective interactions between emergent quasiparticles which constitute a central ingredient of the Landau Fermi liquid theory (LFLT)~\cite{Landau1957FermiLiquid,Landau1957Oscillations,Nozieres1964FermiSystems,PinesNozieres1966QuantumLiquids}. However, extending this description into the
strongly correlated regime in real experiments is nontrivial, since Landau theory is
fundamentally an asymptotic low-energy framework and itself is not restricted to weak microscopic coupling; yet, the challenge is to determine the corresponding energy-functional coefficient nonperturbatively in the strongly interacting regime.

\begin{figure*}[t]
    \centering
    \includegraphics[width=\linewidth]{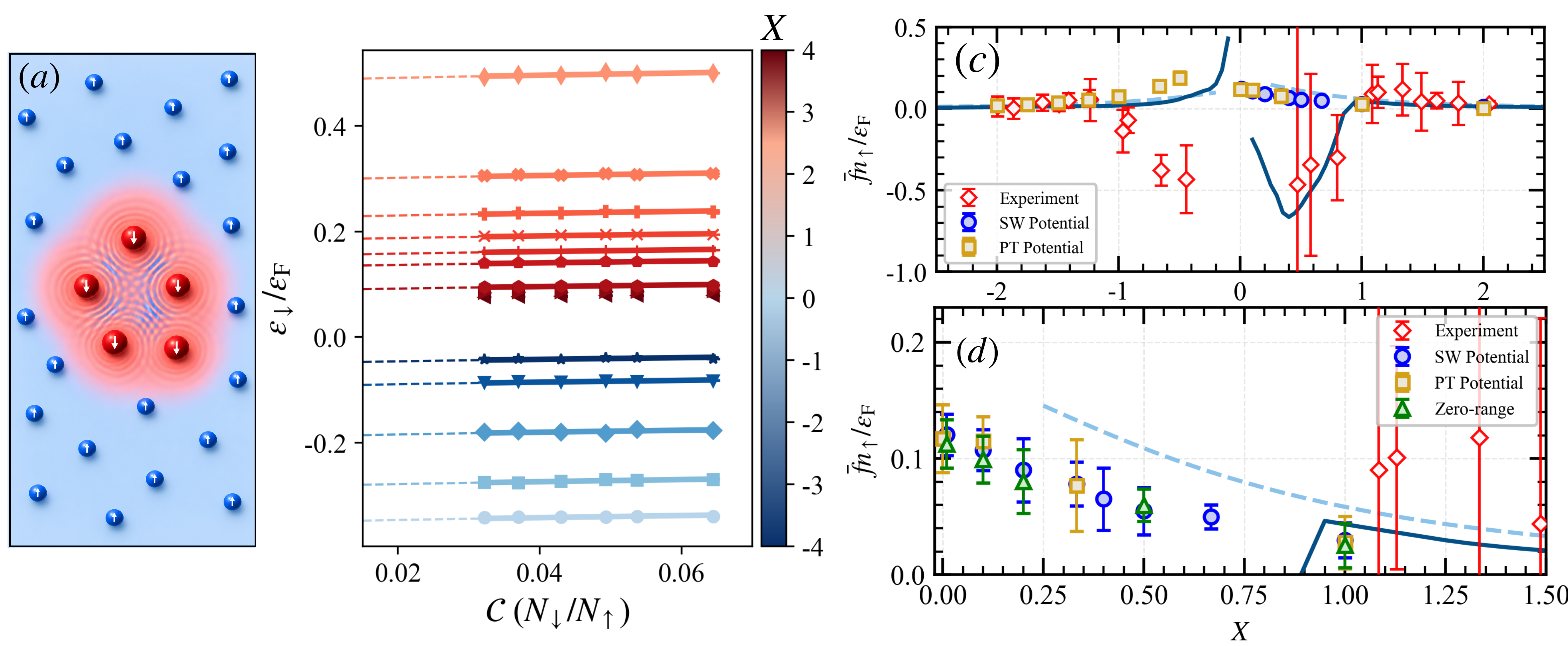}
    \caption{a) Heavy fermionic impurities (red spheres, $\downarrow$) immersed in a Fermi sea of majority atoms (blue small spheres,$\uparrow$), which mediates an effective interaction between the impurities (wiggle lines) and also experience a degeneracy repulsion (red shadow).
(b) Polaron energy versus impurity concentration $\mathcal{C}=N_{\downarrow}/N_{\uparrow}$ for different couplings $X=-(k_{\mathrm F}a)^{-1}$. Fits of the QMC data using Eq.~\ref{eq:EnergyLFLT} yield the single polaron energy $\varepsilon_\downarrow^0$, and the Landau parameter $\bar f n_\uparrow$. The dashed lines indicate the extrapolation to the limit $\mathcal{C} \to 0$.
(c) Induced interaction computed with QMC using different model potentials: A square-well potential (SW), the Bethe–Peierls zero-range boundary condition (BPbc), and Pöschl–Teller potential (PT) for different coupling strengths $X$, against experimental measurements~\cite {bar24} along with both static and dynamical Landau Fermi-liquid predictions.
(d) Strong-coupling regime, showing the persistent repulsive interaction for zero- and finite-range potentials.}
    \label{Fig:Fig1}
\end{figure*}

Ultracold quantum gases provide a versatile platform for studying a broad range of strongly correlated phenomena, including Fermi polarons — quasiparticles formed when neutral impurities are dressed by excitations of the surrounding Fermi gas~\cite{Schirotzek2009,
Kohstall2012,
Koschorreck2012,
Nascimbene2009,
Scazza2022,
Massignan2014,
Chevy2006,
Prokofev2008,
Mora2009,
Combescot2009,
Sidler2017,
Vivanco2025}. However, the polaron concept extends beyond neutral ultracold gases, with related phenomena also studied for charged impurities~\cite{bas21,Perez2021,ast21,wec21,chr22,din22,mysliwy2023longrange,Pessoa2024} and in solid-state platforms~\cite{Tan2020InteractingPolaronPolaritons,Tan2023BosePolaronInteractions}. The properties of a single polaron and its nonequilibrium dynamics are relatively well characterized, whereas interactions between polarons remain much less understood. In highly imbalanced atomic mixtures, medium-induced interactions can give rise to polaron–polaron correlations and even support bipolaron bound states~\cite{CamachoGuardian2018Bipolarons,CamachoGuardian2018Landau,Paredes2024}.  Most recently, experiments with ultracold fermionic mixtures have directly probed interactions between Fermi polarons from weak to strong coupling~\cite{bar24}. For fermionic impurities, the measured interaction agrees with the repulsive static Fermi-liquid prediction at weak and intermediate coupling, yet develops a pronounced downward trend near resonance. Finite-temperature effects, finite-range corrections, and even the preparation protocol are possible explanations for this behavior, leaving the intrinsic nature of the ground-state induced interaction as an important unresolved question.  \\

In this work, we address this question using \textit{ab initio} quantum Monte Carlo methods at zero temperature, specifically tailored to capture all possible virtual particle-hole excitations, strong many-body correlations, and nonlinear density rearrangement in the neighborhood of the impurity~\cite{
Astrakharchik2023}. Our main result is that the Landau interaction between fermionic polarons remains \textit{repulsive} from weak to strong coupling throughout the attractive polaron branch, including near resonance and deep in the universal short-range regime, as systematically verified using several potential models. The repulsive sign of the equilibrium ground-state Landau interaction therefore survives into the strongly correlated regime and is robust against microscopic interaction details, implying that the experimentally observed downturn cannot originate from intrinsic zero-temperature polaron correlations solely. Furthermore, we resolve the corresponding real-space mediated interaction between two static impurities, recovering the  Ruderman–Kittel–Kasuya–Yosida trend at weak coupling and the nonperturbative Born–Oppenheimer (BO) limit at strong coupling~\cite{ens20}. Together, these results provide a microscopic ground-state benchmark for the induced interaction. Our zero-temperature ground-state calculation excludes contributions associated with thermally populated processes, such as Bose-stimulated molecular activation, thereby isolating the equilibrium polaronic contribution to the induced interaction.\\

\textit{System --} Following the recent experimental realization~\cite{bar24}, we consider a highly imbalanced 
two-component Fermi mixture consisting of a majority species of 
$^{6}\mathrm{Li}$ atoms, denoted $\ket{\uparrow}$ and hereafter referred 
to as the bath, doped with a dilute concentration of \textit{degenerate} $^{40}\mathrm{K}$ fermionic 
impurities, denoted $\ket{\downarrow}$ [Fig.~\ref{Fig:Fig1}(a)]. The Hamiltonian 
of the system is given by

\begin{equation}
\mathcal{H} = 
   -\frac{\hbar^2}{2 m_\downarrow} \sum_{j=1}^{N_\downarrow} \nabla_j^2
     -\frac{\hbar^2}{2 m_\uparrow} \sum_{i=1}^{N_\uparrow} \nabla_i^2
    + \sum_{i=1}^{N_\uparrow} \sum_{j=1}^{N_\downarrow} V_{\uparrow\downarrow}(|\mathbf{r}_i - \mathbf{r}_j|),
\label{eq:h}
\end{equation}
where the first two terms represent the kinetic energies of the $N_\downarrow$ impurities (with mass $m_\downarrow$) and the $N_\uparrow$ particles of the fermionic bath (with mass $m_\uparrow$), respectively. Because of Fermi statistics, $s$-wave scattering between identical fermions is forbidden, while higher partial waves are negligible at low temperature. We therefore include interactions only between the two species. These interactions are described by the potential $V(r_{ij})$, where $r_{ij} = |\mathbf{r}_i - \mathbf{r}_j|$ denotes the interparticle distance. The interaction strength is parametrized by $X = -1/(k_{\mathrm{F}} a)$, where $a$ is the scattering length, and  $k_{\mathrm{F}}^{-1} = (6\pi^2 n_{\uparrow})^{-1/3}$  and $\varepsilon_F = \hbar^2 k_F^2/(2m_{\uparrow})$ are the corresponding natural length and energy scales of the bath. In this work, we use the experimental value for the mass ratio $m_\downarrow/m_\uparrow= 6.8$. \\

\textit{Landau quasiparticle interaction --} When multiple impurities are immersed in a Fermi sea, they become dressed by excitations of the surrounding medium, forming polarons that locally perturb the background. These polarons can exchange density fluctuations mediated by the Fermi sea, giving rise to effective interactions between them. Within Landau Fermi-liquid theory, the  quasiparticle energy changes under a variation of the distribution function
\begin{equation}
\bar{\varepsilon}_{\mathbf{p}\sigma}
=
\varepsilon_{\mathbf{p}\sigma}
+
\sum_{\mathbf{k}'\sigma'}
f_{\mathbf{p}\sigma,\mathbf{k}'\sigma'}
\,\delta n_{\mathbf{k}'\sigma'},
\end{equation}
where $f_{\mathbf{p}\sigma,\mathbf{k}'\sigma'}$ is the Landau quasiparticle interaction parameter and $\delta n_{\mathbf{k}'\sigma'}$ denotes the change in the quasiparticle occupation numbers with respect to the reference state. For the minority component, we characterize the induced interaction by
the momentum-averaged Landau parameter $\bar f$.  At zero temperature,
the many-polaron energy can be obtained from the Landau-Pomeranchuk expansion

\begin{equation}
\varepsilon_\downarrow
=
\varepsilon_\downarrow^0
+
\frac{3}{5} \varepsilon_F \frac{m_\uparrow}{m^*} \mathcal{C}^{2/3}
+
\frac{1}{2}\bar f\, n_\uparrow \mathcal{C}
+\cdots
\label{eq:EnergyLFLT}
 \end{equation}
 where $\varepsilon_\downarrow^0$ is the single-polaron energy, the second term represents the Fermi pressure depending on the effective mass, and we define the concentration as $\mathcal{C}=N_{\downarrow}/N_{\uparrow}\ll 1$. In order to extract $\bar f$, we use six impurity concentrations in the range $0.03\leq \mathcal C\leq 0.07$ and compute the many-polaron energy using QMC as $\varepsilon_\downarrow = \bigl[E(N_\uparrow,N_\downarrow)-E(N_\uparrow,0)\bigr]/N_\downarrow$. Additional calculations extending to $\mathcal C \sim 0.15$ show no statistically significant deviation from the fitted behavior within our numerical uncertainties. On the other hand, the experiment of Ref.~\cite{bar24} extracts the corresponding momentum-averaged interaction from RF spectroscopy by fitting its concentration dependence to a linear form. In this case, the Fermi-pressure contribution is absent because the impurity component remains nondegenerate. This constitutes a fundamental difference between the experimental system and our zero-temperature calculation with degenerate fermionic impurities. \\

 In Fig.~\ref{Fig:Fig1}(b) we report the many - polaron energy  $\varepsilon_{\downarrow}(C)$ as a function of the concentration for different values of the coupling strength $X$, for both the attractive (blue shading) and repulsive (red shading) branches. both the single-polaron energy  $\varepsilon_\downarrow^0$ and the induced interaction parameter $\bar f n_\uparrow$ are obtained by fitting according to Eq.~(\ref{eq:EnergyLFLT}). Interestingly, for a degenerate impurity Fermi gas, the Landau–Pomeranchuk theory predicts a Fermi-pressure contribution proportional to $\mathcal{C}^{2/3}$, which can be comparable in magnitude or larger with respect to the Landau interaction term according to our simulations. However, the two contributions have distinct physical origins. The Fermi pressure reflects the kinetic-energy cost of filling the impurity Fermi sea and is present even for noninteracting quasiparticles, whereas the Landau parameter encodes the effective interaction between quasiparticles.\\

\textit{Mediated Interaction --} Our central result is that the  Landau interaction for fermionic impurities remains \textit{positive} throughout the attractive polaronic branch, including the strongly interacting regime near the resonance. In Fig.~\ref{Fig:Fig1}(c), we show the induced interaction $\bar f n_{\uparrow}/\varepsilon_F$ as a function of the coupling strength $X$. Our calculations are done for both the attractive $X>0$ and for the repulsive $X<0$ branch using  QMC calculations within the fixed-node approximation. In the weak-coupling regime, $|X|\gg1$, our QMC results quantitatively agree with theoretical predictions from Landau Fermi-liquid theory (LFLT) for both static and dynamical properties, as well as with experimental measurements on both polaron branches~\cite{bar24}.
This agreement persists into the intermediate-coupling regime,
$|X|\sim1$, within the numerical and experimental uncertainties. These findings are consistent with the expected repulsive interaction
for fermionic degenerate impurities in the fixed-density static limit ~\cite{YuPethick2012,GiraudCombescot2012} originating from the Pauli-blocking-induced sign of the exchange contribution.\\

\begin{figure}[t] 
    \centering
    \includegraphics[width=\columnwidth]{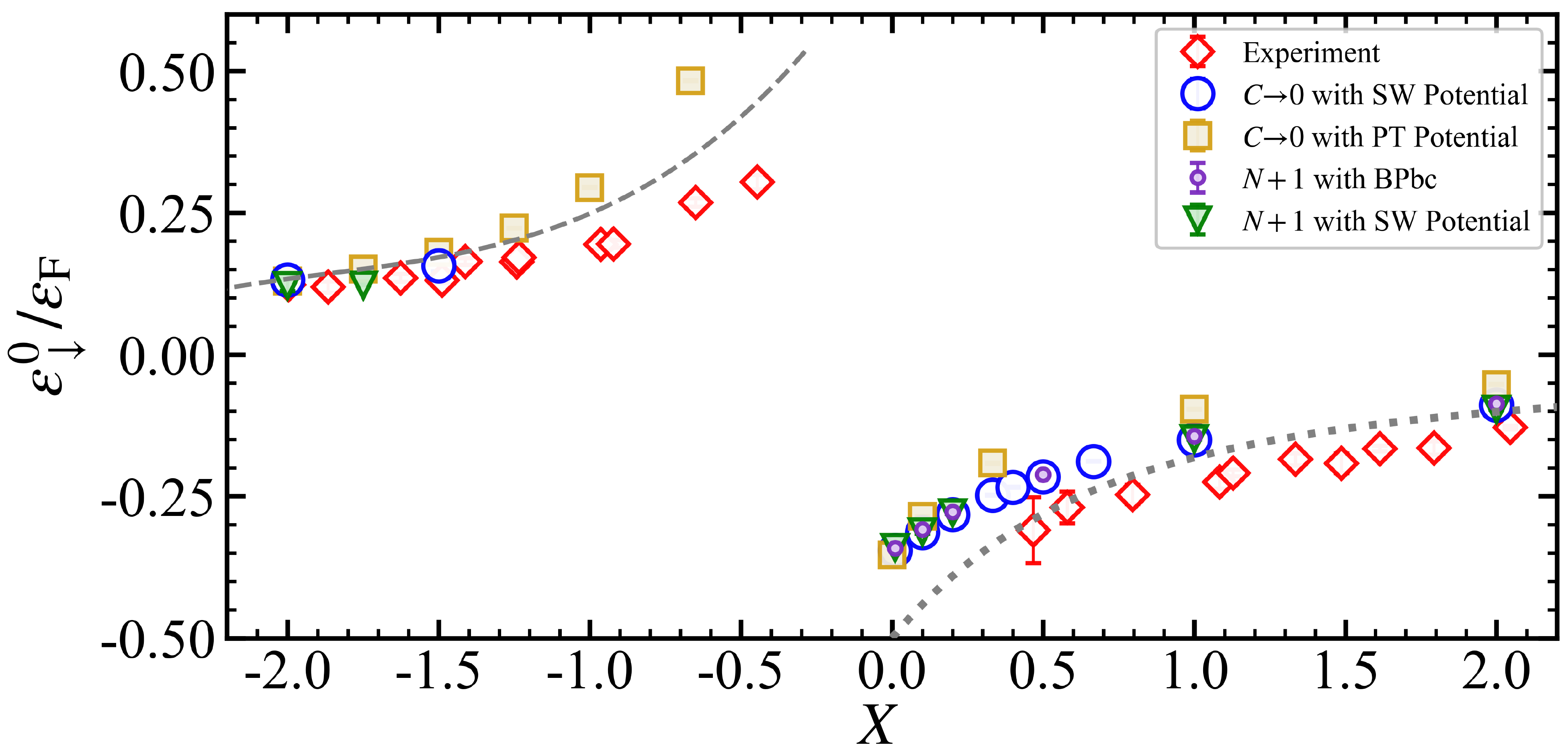}
    \caption{Single polaron energy as a function of the coupling strength
$X$. QMC results in the $\mathcal C \to0$ limit are compared with direct $N+1$
calculations, experimental measurements, and many-body theoretical predictions. The infinite-mass limit is plotted for the attractive branch (Dashed line), whereas the repulsive branch is described by the perturbative prediction of Ref. \cite{Massignan2011Repulsive} (dotted line).  }
\label{Fig:Fig2}
\end{figure}

In the strongly interacting regime and near resonance, the behavior changes qualitatively. The experimentally measured mediated interaction exhibits a pronounced downturn toward negative values, whereas our zero-temperature QMC results remain positive within the polaronic branch (attractive branch), as shown in Fig.~\ref{Fig:Fig1}(c). Our QMC projects onto the lowest-energy state compatible with the imposed polaronic nodal surface, defined by plane-wave Slater determinants. This constraint suppresses projection onto the molecular ground state, whose localized bound-state character generally entails a distinct nodal structure, thereby stabilizing the metastable repulsive-polaron branch. To assess the universality of our results, we performed QMC calculations using three different interaction models, namely the zero-range Bethe--Peierls boundary condition (BPbc), a finite-range Pöschl -- Teller (PT), and a square-well (SW) potential chosen such that $nR_0^3=10^{-6}$ (see End Matter). All calculations lie well within the universal short-range regime,
 within statistical uncertainties, show no significant dependence
on the interaction model. The results remain consistent with
$\bar f>0$, as shown in Fig.~\ref{Fig:Fig1}(d). On the other hand, the experiment, by comparison, involves a finite resonance-range parameter, a finite temperature bath $T/T_\mathrm{F} \sim 0.25
$ and a nonequilibrium preparation protocol. The measured downturn cannot be accounted for by equilibrium zero-temperature polaron correlations alone.  The contribution from the latter is positive and is extracted from the \textit{ab initio} calculation that isolates the intrinsic zero-temperature polaronic contribution. The persistence of a positive $\bar{f}$ therefore suggests that strong polaron correlations alone cannot account for the experimental downturn.

\begin{figure*}[t]
    \centering
    \includegraphics[width=\textwidth]{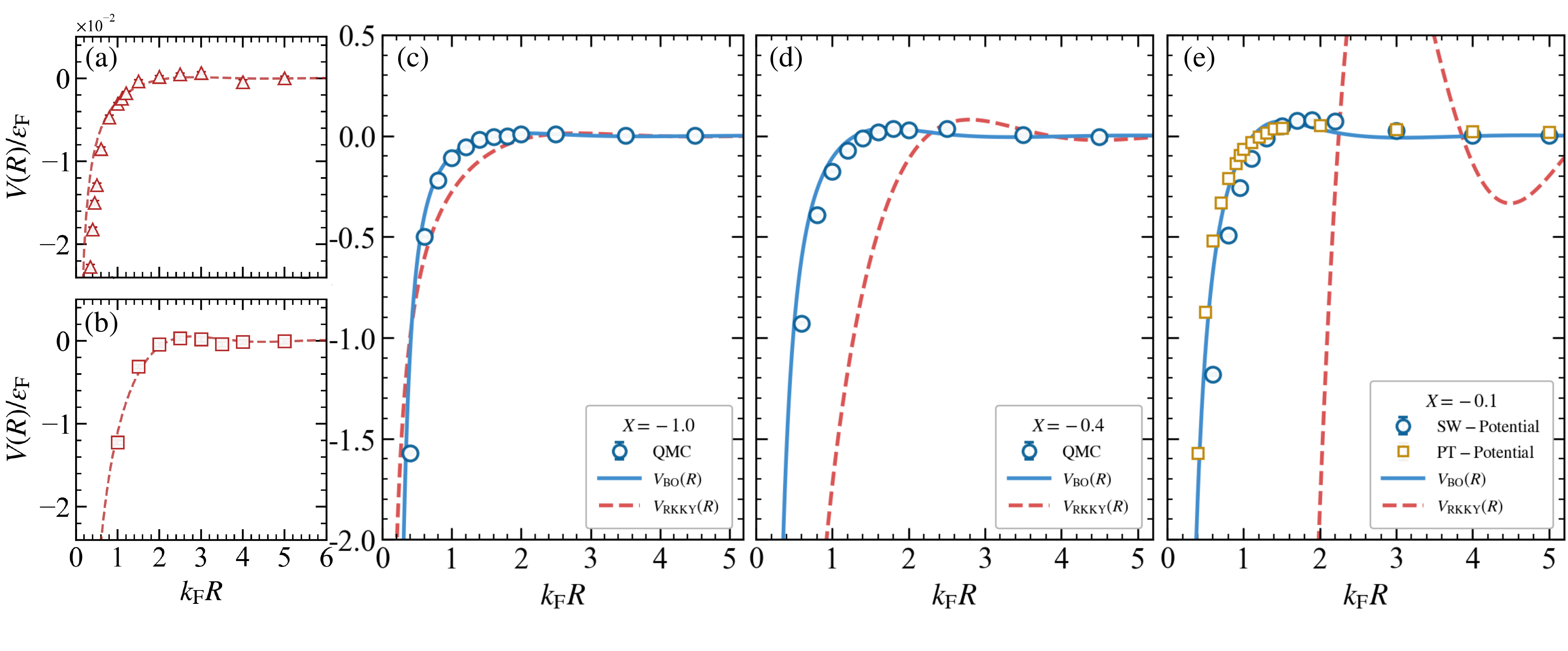}
    \caption{
        Induced potential in the weakly interacting regime for
        (a) \(X=-10\) and
        (b) \(X=-5\), and in the strongly interacting regime for
        (c) \(X=-1\),
        (d) \(X=-0.25\), and
        (e) \(X=-0.1\), using both a SW and a PT potential. In all panels, the red dashed lines represent the RKKY prediction.
    }
    \label{FIG:induced_potential}
\end{figure*}

The sign of the mediated interaction can depend on the thermodynamic constraint under which the response is evaluated, in particular whether the bath density or chemical potential is held fixed as recently investigated~\cite{LevinsenParish2025}. In injection RF spectroscopy, impurities are transferred from a weakly interacting to an interacting state on a timescale shorter than that required for global equilibration of the bath. The measurement therefore probes a predominantly local response, for which a fixed-density description provides the relevant picture, consistent with the measurement in our numerical methodology. After separating the kinetic contribution (Fermi pressure) from the interaction term, our QMC results provide the corresponding zero-temperature fixed-density benchmark for the mediated interaction. Residual differences with experiment may therefore originate from finite temperature and molecular correlations, finite resonance range and the nonequilibrium preparation protocol.\\

\textit{Single Polaron Energy —}  We extract the polaron energy from the constant term in Eq.~\ref{eq:EnergyLFLT}. In Fig.~\ref{Fig:Fig2}, we report the values obtained directly from the fittings from the curves in Fig.~\ref{Fig:Fig1}(b). We compare these results with direct calculations of the $N+1$ polaron problem, corresponding to the independent calculation of the impurity chemical potential~\cite{lob06ns,vli13,pes21}. The $N+1$ calculations are performed both with the same finite-range interaction (SW and PT potentials) and with the strictly zero-range limit, implemented through the BPbc. The agreement between the two independent calculations validates the energy coefficient independently and indicates that finite-range
corrections are negligible within our numerical uncertainties. On the attractive branch, for $0<X\lesssim 0.5$, our QMC energies lie slightly above the experimental values. However, our calculations are performed
deep into the universal short-range regime and $T=0$, whereas the experiment has a finite resonance-range parameter. On the repulsive branch, the agreement is reasonably good in the weak- and intermediate-coupling regimes, while at strong coupling our QMC results qualitatively reproduce the experimental trend. Of course, the energy is also sensitive to finite-range effects and finite-temperature fluctuations, which may constitute possible sources of the residual discrepancy, particularly in the strong-coupling regime.~\cite{pes21}. \\

\textit{Real-space Mediated Interaction—} While the Landau parameter $\bar{f}$ characterizes the change in the energy of mobile quasiparticle excitations with their density, it is also important to characterize the spatial structure of the underlying interaction mechanism. To this end, we consider the complementary static-impurity limit. Specifically, we consider two fixed impurities and determine the medium-induced interaction potential as a function of their separation $R$. Previous diagrammatic Monte Carlo calculations show that the attractive-polaron
energy varies only weakly with impurity mass once the impurity is sufficiently
heavy~\cite{Kroiss2015}. Consistently, our Fig.~\ref{Fig:Fig2} places the experimental
mass ratio $m_\downarrow/m_\uparrow=6.8$ close to the static-impurity limit. Motivated by this observation,
we investigate the medium-induced interaction potential between
two static impurities. Its oscillatory spatial dependence is
analogous to that of the Ruderman--Kittel--Kasuya--Yosida (RKKY)~\cite{RudermanKittel1954}
interaction between localized magnetic moments mediated by
conduction electrons.\\

To obtain the spatially resolved induced interaction in a controlled setting, we separately consider the static-impurity limit
$m_{\downarrow}=10^{10}m_{\uparrow}$ and fix the two impurities at a separation $R$. The induced interaction potential is obtained
from the ground-state energies as
\begin{equation}
V_{\mathrm{DMC}}(R)
=
E_{N+2}^{\mathrm{DMC}}(R)
-2E_{N+1}^{\mathrm{DMC}}
+E_N^{\mathrm{DMC}},
\end{equation}
where $E_N^{\mathrm{DMC}}$ is the energy of the bath containing
$N$ majority fermions, $E_{N+1}^{\mathrm{DMC}}$ is the energy
of the bath with one impurity, and $E_{N+2}^{\mathrm{DMC}}(R)$
is the energy of the bath with two impurities separated by $R$.
All three energies are evaluated for the same bath particle
number and volume.

In the weak-coupling regime, we calculate the induced potential
at $X=-10$ and $X=-5$, and
compare our results with the perturbative RKKY-like expression~\cite{RudermanKittel1954,ens20,Ando2025Mediated},
\begin{equation}
V_{\mathrm{RKKY}}(R)
=
\frac{\varepsilon_{\mathrm F}}{2\pi X^{2}}
\frac{
(2k_{\mathrm F}R)\cos(2k_{\mathrm F}R)
-\sin(2k_{\mathrm F}R)
}{
(k_{\mathrm F}R)^{4}
},
\label{eq:RKKY}
\end{equation}

Our results agree with both the standard Born-Oppenheimer approximation and the perturbative RKKY expression in Eq.~(\ref{eq:RKKY}), as shown in Fig.~\ref{FIG:induced_potential}(a-b)\\

For the strongly interacting regime, we use a nonperturbative approach based on the Born–Oppenheimer treatment~\cite{ens20}, valid for arbitrary coupling in the static-impurity limit. Within this approach, the induced potential is obtained from the bound-state energies and scattering phase shifts of bath fermions interacting with two fixed scatterers through zero-range interactions. Subtracting twice the isolated-impurity energy yields $V(R)=\Delta E(R)-2\mu$, providing a direct benchmark for our DMC energy differences in the zero-range and thermodynamic limits.  Fig.~\ref{FIG:induced_potential}(c-e) shows excellent agreement between the BO and DMC results over all separations studied, including the small oscillations characteristic of the RKKY-like long-distance behavior. At our strongest interaction strength, $X=-0.1$, calculations with a Pöschl–Teller (PT) potential yield consistent results, further supporting the universal character of the interaction. In this regime, the oscillation amplitude is much smaller than that predicted by the weak-coupling perturbative expression. At weak coupling, the impurity induces only a small density modulation around a homogeneous Fermi sea, and the induced interaction can be obtained from the RKKY theory. At strong coupling, the density around each impurity becomes strongly inhomogeneous and nonlinear~\cite{Pessoa2024}, so the local homogeneous-density picture breaks down near the impurity and the induced interaction must be obtained from the fully rearranged many-body state. 
 In addition, our calculations establish that the positive induced interaction is not an artifact of either finite system size or the particular finite-range interaction model. Thus, its persistence at strong coupling is consistent with being an intrinsic ground-state many-body effect.\\
 
  The attractive regions of the static-impurity potential do not contradict the positive Landau interaction for mobile fermionic polarons. These quantities probe different limits, on the one hand $V(R)$ describes fixed scatterers at a fixed distance; on the other hand, $\bar f$ characterizes the response of mobile quasiparticles, where Fermi statistics and Pauli exchange play an essential role.\\

\textit{Conclusions--} By means of \textit{ab initio} QMC methods at zero temperature, we show that the Landau interaction between fermionic quasiparticles remains positive throughout the attractive branch. Although the repulsive branch also displays a repulsive Landau interaction, this is not a ground-state calculation. In the polaronic branch, the sign is robust against the choice of microscopic interaction model, demonstrating that strong polaron correlations alone do not reproduce the downturn observed experimentally in the strongly correlated regime near resonance. Our results therefore provide a microscopic reference for separating intrinsic polaron interactions from finite-temperature effects, molecular correlations, finite-range effects, and preparation-dependent effects. As a complementary probe of the same mediation mechanism, we determine the real-space interaction between two static impurities. Our calculations reproduce the nonperturbative Born--Oppenheimer potential, including its attractive regions and weak long-distance oscillations. This agreement establishes a quantitative benchmark beyond the perturbative RKKY regime and provides a complementary benchmark to the Landau interaction obtained for mobile fermionic quasiparticles.

These results provide microscopic benchmarks for theories of strongly interacting quasiparticles and establish a zero-temperature framework for disentangling equilibrium polaron interactions from thermal, molecular, finite-range, and preparation-dependent effects in experiments. In this work we explore the case of fermionic impurities, yet an interesting question also aligned with experiments is to explore bosonic systems~\cite {Ardila2022UltraDilute,Grusdt2025Impurities,Levinsen2025MediumEnhanced}. Beyond ultracold atoms, they address the broader question of how a fermionic environment mediates interactions between dressed particles, a problem also relevant to exciton polarons in doped semiconductors and carrier-mediated interactions in metals~\cite{Massignan2026Polarons}.\\

\textit{Acknowledgements --} We thank Tilman Enss, Jesper Levinsen, Cosetta Baroni, Shibalik Lahiri and Valentin Walther for insightful discussions. S.A.V. acknowledges financial support from the Fundação de Amparo à Pesquisa do Estado de São Paulo (FAPESP; São Paulo Research Foundation), grant No.~2023/07225-0.. Computational resources have been provided in part by CENAPAD-SP at Unicamp. R.P. thanks the computer support from LaMCAD/UFG.

\bibliographystyle{apsrev4-2}
\bibliography{pessoa}

@article{bar24,
  author       = {Cosetta Baroni and Bo Huang and Isabella Fritsche and Erich Dobler and Gregor Anich and Emil Kirilov and Rudolf Grimm and Miguel A. Bastarrachea‑Magnani and Pietro Massignan and Georg M. Bruun},
  title        = {Mediated interactions between Fermi polarons and the role of impurity quantum statistics},
  journal      = {Nature Physics},
  year         = {2024},
  volume       = {20},
  pages        = {68--73},
  doi          = {10.1038/s41567-023-02248-4},
}

@article{
LevinsenParish2025,
  title   = {Role of impurity statistics and medium constraints in polaron-polaron interactions},
  author  = {Levinsen, Jesper and Marchetti, Francesca M. and Bleu, Olivier and Parish, Meera M.},
  journal = {SciPost Phys.},
  volume  = {21},
  pages   = {025},
  year    = {2026},
  doi     = {10.21468/SciPostPhys.21.1.025}
}

@Article{wec21,
  author       = {Pascal Weckesser and Fabian Thielemann and Dariusz Wiater and Agata Wojciechowska and Leon Karpa and Krzysztof Jachymski and Micha{\l} Tomza and Thomas Walker and Tobias Schaetz},
  journaltitle = {Nature},
  title        = {Observation of Feshbach resonances between a single ion and ultracold atoms},
  doi          = {10.1038/s41586-021-04112-y},
  number       = {7889},
  pages        = {429--433},
  volume       = {600},
  journal      = {Nature},
  publisher    = {Springer Science and Business Media {LLC}},
  year         = {2021}
}

@article{Kroiss2015,
  author  = {Kroi{\ss}, Peter and Pollet, Lode},
  title   = {Diagrammatic Monte Carlo study of a mass-imbalanced {Fermi}-polaron system},
  journal = {Physical Review B},
  volume  = {91},
  pages   = {144507},
  year    = {2015},
  doi     = {10.1103/PhysRevB.91.144507}
}

@article{Massignan2011Repulsive,
  author  = {Massignan, P. and Bruun, G. M.},
  title   = {Repulsive polarons and itinerant ferromagnetism in strongly polarized Fermi gases},
  journal = {The European Physical Journal D},
  volume  = {65},
  pages   = {83--89},
  year    = {2011},
  doi     = {10.1140/epjd/e2011-20084-5}
}

@article{Berk1966,
  author  = {Berk, N. F. and Schrieffer, J. R.},
  title   = {Effect of Ferromagnetic Spin Correlations on Superconductivity},
  journal = {Physical Review Letters},
  volume  = {17},
  pages   = {433--435},
  year    = {1966},
  doi     = {10.1103/PhysRevLett.17.433}
}

@article{CamachoGuardian2018Bipolarons,
  author  = {Camacho-Guardian, A. and Pe{\~n}a Ardila, L. A. and Pohl, T. and Bruun, G. M.},
  title   = {Bipolarons in a Bose-Einstein Condensate},
  journal = {Physical Review Letters},
  volume  = {121},
  number  = {1},
  pages   = {013401},
  year    = {2018},
  doi     = {10.1103/PhysRevLett.121.013401}
}

@article{CamachoGuardian2018Landau,
  author  = {Camacho-Guardian, A. and Bruun, G. M.},
  title   = {Landau Effective Interaction between Quasiparticles in a Bose-Einstein Condensate},
  journal = {Physical Review X},
  volume  = {8},
  number  = {3},
  pages   = {031042},
  year    = {2018},
  doi     = {10.1103/PhysRevX.8.031042}
}

@article{Tan2020InteractingPolaronPolaritons,
  author  = {Tan, Li Bing and Cotlet, Ovidiu and Bergschneider, Andrea
             and Schmidt, Richard and Back, Patrick and Shimazaki, Yuya
             and Kroner, Martin and Imamoglu, Atac},
  title   = {Interacting Polaron-Polaritons},
  journal = {Physical Review X},
  volume  = {10},
  number  = {2},
  pages   = {021011},
  year    = {2020},
  doi     = {10.1103/PhysRevX.10.021011}
}

@article{Tan2023BosePolaronInteractions,
  author  = {Tan, Li Bing and Diessel, Oriana K. and Popert, Alexander
             and Schmidt, Richard and Imamoglu, Atac and Kroner, Martin},
  title   = {Bose Polaron Interactions in a Cavity-Coupled Monolayer Semiconductor},
  journal = {Physical Review X},
  volume  = {13},
  number  = {3},
  pages   = {031036},
  year    = {2023},
  doi     = {10.1103/PhysRevX.13.031036}
}

@article{Chevy2006,
  title = {Universal phase diagram of a strongly interacting Fermi gas with unbalanced spin populations},
  author = {Chevy, F.},
  journal = {Phys. Rev. A},
  volume = {74},
  issue = {6},
  pages = {063628},
  numpages = {4},
  year = {2006},
  month = {Dec},
  publisher = {American Physical Society},
  doi = {10.1103/PhysRevA.74.063628},
  url = {https://link.aps.org/doi/10.1103/PhysRevA.74.063628}
}

@article{Scazza2022,
  title         = {Repulsive Fermi and Bose Polarons in Quantum Gases},
  author        = {Francesco Scazza and Matteo Zaccanti and Pietro Massignan and Meera M. Parish and Jesper Levinsen},
  journal       = {Atoms},
  year          = {2022},
  volume        = {10},
  number        = {2},
  article-number = {55},
  doi           = {10.3390/atoms10020055},
  issn          = {2218-2004},
  publisher     = {MDPI AG}
}

@article{ard15,
  title = {Impurity in a Bose-Einstein condensate: Study of the attractive and repulsive branch using quantum Monte Carlo methods},
  author = {L. A. Pe{\~{n}}a Ardila and S. Giorgini},
  journal = {Phys. Rev. A},
  volume = {92},
  issue = {3},
  pages = {033612},
  numpages = {12},
  year = {2015},
  month = {Sep},
  publisher = {American Physical Society},
  doi = {10.1103/PhysRevA.92.033612},
  url = {https://link.aps.org/doi/10.1103/PhysRevA.92.033612}
}

@Article{ast04,
  author    = {Astrakharchik, G. E. and Boronat, J. and Casulleras, J. and Giorgini, S.},
  title     = {Equation of State of a {F}ermi Gas in the BEC-BCS Crossover: A Quantum Monte Carlo Study},
  doi       = {10.1103/PhysRevLett.93.200404},
  issue     = {20},
  pages     = {200404},
  url       = {http://link.aps.org/doi/10.1103/PhysRevLett.93.200404},
  volume    = {93},
  journal   = {Phys. Rev. Lett.},
  month     = {Nov},
  numpages  = {4},
  publisher = {American Physical Society},
  year      = {2004}
}

@Article{ast21,
  author    = {Grigory E. Astrakharchik and Luis A. {Pe{\~{n}}a Ardila} and Richard Schmidt and Krzysztof Jachymski and Antonio Negretti},
  title     = {Ionic polaron in a Bose-Einstein condensate},
  doi       = {10.1038/s42005-021-00597-1},
  year      = {2021},
  number    = {1},
  volume    = {4},
  pages     = {94},
  journal   = {Commun. Phys.},
  publisher = {Springer Science and Business Media {LLC}}
}

@article{YuPethick2012,
  author  = {Yu, Zhenhua and Pethick, Christopher J.},
  title   = {Induced interactions in dilute atomic gases and liquid helium mixtures},
  journal = {Physical Review A},
  volume  = {85},
  number  = {6},
  pages   = {063616},
  year    = {2012},
  doi     = {10.1103/PhysRevA.85.063616}
}

@article{GiraudCombescot2012,
  author  = {Giraud, S. and Combescot, R.},
  title   = {Interaction between polarons and analogous effects in polarized Fermi gases},
  journal = {Physical Review A},
  volume  = {85},
  number  = {1},
  pages   = {013605},
  year    = {2012},
  doi     = {10.1103/PhysRevA.85.013605}
}

@article{bas21,
  author  = {Bastarrachea-Magnani, Miguel A. and Camacho-Guardian, Arturo and Bruun, Georg M.},
  title   = {Attractive and Repulsive Exciton-Polariton Interactions Mediated by an Electron Gas},
  journal = {Phys. Rev. Lett.},
  volume  = {126},
  number  = {12},
  pages   = {127405},
  year    = {2021},
  doi     = {10.1103/PhysRevLett.126.127405}
}

@Article{car03,
  author    = {J. Carlson and S.-Y. Chang and V. R. Pandharipande and K. E. Schmidt},
  title     = {Superfluid {F}ermi Gases with Large Scattering Length},
  eid       = {050401},
  number    = {5},
  pages     = {050401},
  url       = {http://link.aps.org/abstract/PRL/v91/e050401},
  volume    = {91},
  journal   = {Phys. Rev. Lett.},
  numpages  = {4},
  publisher = {APS},
  year      = {2003},
}

@Article{car05,
  author    = {Carlson, J. and Reddy, Sanjay},
  title     = {Asymmetric Two-Component Fermion Systems in Strong Coupling},
  doi       = {10.1103/PhysRevLett.95.060401},
  issue     = {6},
  pages     = {060401},
  url       = {http://link.aps.org/doi/10.1103/PhysRevLett.95.060401},
  volume    = {95},
  journal   = {Phys. Rev. Lett.},
  month     = {Aug},
  numpages  = {4},
  publisher = {American Physical Society},
  year      = {2005},
}

@Article{cha04,
  author    = {S. Y. Chang and V. R. Pandharipande and J. Carlson and K. E. Schmidt},
  title     = {Quantum {M}onte {C}arlo studies of superfluid {F}ermi gases},
  eid       = {043602},
  number    = {4},
  pages     = {043602},
  url       = {http://link.aps.org/abstract/PRA/v70/e043602},
  volume    = {70},
  journal   = {Physical Review A (Atomic, Molecular, and Optical Physics)},
  numpages  = {11},
  publisher = {APS},
  year      = {2004},
}

@Article{cha11,
  Title                    = {{Ferromagnetism in the upper branch of the Feshbach resonance and the hard-sphere {F}ermi gas}},
  Author                   = {{Chang}, S.-Y. and {Randeria}, M. and {Trivedi}, N.},
  Journal                  = {Proceedings of the National Academy of Science},
  Year                     = {2011},

  Month                    = Jan,
  Pages                    = {51-54},
  Volume                   = {108},
  Adsurl                   = {http://adsabs.harvard.edu/abs/2011PNAS..108...51C},
  Doi                      = {10.1073/pnas.1011990108}
}

@article{chr22,
  title = {Mobile ion in a Fermi sea},
  author = {Christensen, Esben Rohan and Camacho-Guardian, Arturo and Bruun, Georg M.},
  journal = {Phys. Rev. A},
  volume = {105},
  issue = {2},
  pages = {023309},
  numpages = {8},
  year = {2022},
  month = {Feb},
  publisher = {American Physical Society},
  doi = {10.1103/PhysRevA.105.023309},
  url = {https://link.aps.org/doi/10.1103/PhysRevA.105.023309}
}

@Article{cow02,
  author    = {{Cowell}, S. and {Heiselberg}, H. and {Mazets}, I.~E. and {Morales}, J. and {Pandharipande}, V.~R. and {Pethick}, C.~J.},
  title     = {{Cold Bose Gases with Large Scattering Lengths}},
  doi       = {10.1103/PhysRevLett.88.210403},
  eid       = {210403},
  eprint    = {cond-mat/0106628},
  number    = {21},
  pages     = {210403},
  volume    = {88},
  adsurl    = {http://adsabs.harvard.edu/abs/2002PhRvL..88u0403C},
  journal   = {Phys. Rev. Lett.},
  month     = May,
  year      = {2002},
}

@article{din22,
  title = {Mediated Interaction between Ions in Quantum Degenerate Gases},
  author = {Ding, Shanshan and Drewsen, Michael and Arlt, Jan J. and Bruun, G. M.},
  journal = {Phys. Rev. Lett.},
  volume = {129},
  issue = {15},
  pages = {153401},
  numpages = {6},
  year = {2022},
  month = {Oct},
  publisher = {American Physical Society},
  doi = {10.1103/PhysRevLett.129.153401},
  url = {https://link.aps.org/doi/10.1103/PhysRevLett.129.153401}
}

@article{ens20,
  author  = {Enss, Tilman and Tran, Binh and Rautenberg, Michael and Gerken, Manuel and Lippi, Eleonora and Drescher, Moritz and Zhu, Bing and Weidem{\"u}ller, Matthias and Salmhofer, Manfred},
  title   = {Scattering of two heavy Fermi polarons: Resonances and quasibound states},
  journal = {Phys. Rev. A},
  volume  = {102},
  number  = {6},
  pages   = {063321},
  year    = {2020},
  doi     = {10.1103/PhysRevA.102.063321}
}

@Article{for11,
  author        = {Forbes, Michael McNeil and Gandolfi, Stefano and Gezerlis, Alexandros},
  title         = {{Resonantly Interacting Fermions in a Box}},
  journal       = {Phys. Rev. Lett.},
  year          = {2011},
  volume        = {106},
  number        = {23},
  pages         = {235303},
  month         = jun,
  adsurl        = {http://adsabs.harvard.edu/abs/2011PhRvL.106w5303F},
  archiveprefix = {arXiv},
  doi           = {10.1103/PhysRevLett.106.235303},
  eid           = {235303},
  eprint        = {1011.2197},
  primaryclass  = {cond-mat.quant-gas},
}

@Article{for12,
  author    = {Michael McNeil Forbes and Stefano Gandolfi and Alexandros Gezerlis},
  title     = {Effective-range dependence of resonantly interacting fermions},
  doi       = {10.1103/physreva.86.053603},
  number    = {5},
  volume    = {86},
  journal   = {Physical Review A},
  month     = {nov},
  publisher = {American Physical Society ({APS})},
  year      = {2012},
}

@Article{gan11,
  author    = {Gandolfi, S. and Schmidt, K. E. and Carlson, J.},
  title     = {BEC-BCS crossover and universal relations in unitary {F}ermi gases},
  doi       = {10.1103/PhysRevA.83.041601},
  issue     = {4},
  pages     = {041601},
  url       = {http://link.aps.org/doi/10.1103/PhysRevA.83.041601},
  volume    = {83},
  journal   = {Phys. Rev. A},
  month     = {Apr},
  numpages  = {4},
  publisher = {American Physical Society},
  year      = {2011},
}

@Article{gan14,
  Title                    = {{Quantum Monte Carlo study of strongly interacting {F}ermi gases}},
  Author                   = {{Gandolfi}, S.},
  Journal                  = {Journal of Physics Conference Series},
  Year                     = {2014},

  Month                    = Aug,
  Number                   = {1},
  Pages                    = {012011},
  Volume                   = {529},
  Adsurl                   = {http://adsabs.harvard.edu/abs/2014JPhCS.529a2011G},
  Archiveprefix            = {arXiv},
  Doi                      = {10.1088/1742-6596/529/1/012011},
  Eid                      = {012011},
  Eprint                   = {1312.1612},
  Primaryclass             = {cond-mat.quant-gas}
}

@Article{gez09,
  author    = {Gezerlis, Alexandros and Gandolfi, S. and Schmidt, K. E. and Carlson, J.},
  title     = {Heavy-Light Fermion Mixtures at Unitarity},
  journal   = {Phys. Rev. Lett.},
  year      = {2009},
  volume    = {103},
  number    = {6},
  pages     = {060403-1},
  month     = {Aug},
  issn      = {1079-7114},
  doi       = {10.1103/physrevlett.103.060403},
  publisher = {American Physical Society (APS)},
  url       = {http://dx.doi.org/10.1103/PhysRevLett.103.060403},
}

@Article{gio08,
  author    = {Giorgini, Stefano and Pitaevskii, Lev P. and Stringari, Sandro},
  title     = {Theory of ultracold atomic {F}ermi gases},
  doi       = {10.1103/RevModPhys.80.1215},
  issue     = {4},
  pages     = {1215--1274},
  url       = {http://link.aps.org/doi/10.1103/RevModPhys.80.1215},
  volume    = {80},
  journal   = {Rev. Mod. Phys.},
  month     = {Oct},
  publisher = {American Physical Society},
  year      = {2008},
}

@Article{lob06,
  author    = {C. Lobo and I. Carusotto and S. Giorgini and A. Recati and S. Stringari},
  title     = {Pair Correlations of an Expanding Superfluid {F}ermi Gas},
  journal   = {Physical Review Letters},
  year      = {2006},
  volume    = {97},
  number    = {10},
  month     = {Sep},
  doi       = {10.1103/physrevlett.97.100405},
  publisher = {American Physical Society ({APS})},
}

@Article{lob06ns,
  author    = {{Lobo}, C. and {Recati}, A. and {Giorgini}, S. and {Stringari}, S.},
  title     = {{Normal State of a Polarized {F}ermi Gas at Unitarity}},
  journal   = {Phys. Rev. Lett.},
  year      = {2006},
  volume    = {97},
  number    = {20},
  pages     = {200403},
  month     = Nov,
  adsurl    = {http://adsabs.harvard.edu/abs/2006PhRvL..97t0403L},
  doi       = {10.1103/PhysRevLett.97.200403},
  eid       = {200403},
  eprint    = {cond-mat/0607730},
}

@article{Pessoa2024,
  author  = {Pessoa, Renato and Vitiello, S. A. and Pe{\~n}a Ardila, L. A.},
  title   = {Fermi Polaron in Atom-Ion Hybrid Systems},
  journal = {Physical Review Letters},
  volume  = {133},
  pages   = {233002},
  year    = {2024},
  doi     = {10.1103/PhysRevLett.133.233002}
}

@Article{pes19,
  author    = {Renato Pessoa and S. A. Vitiello and K. E. Schmidt},
  title     = {Zero-range {F}ermi gas along the {BCS}-{BEC} crossover},
  doi       = {10.1103/physreva.100.053601},
  number    = {5},
  volume    = {100},
  journal   = {Physical Review A},
  month     = {Nov},
  publisher = {American Physical Society ({APS})},
  year      = {2019},
}

@Article{Astrakharchik2023,
author={Astrakharchik, Grigory E.
and Pe{\~{n}}a Ardila, Luis A. 
and Jachymski, Krzysztof
and Negretti, Antonio},
title={Many-body bound states and induced interactions of charged impurities in a bosonic bath},
journal={Nat. Commun.},
year={2023},
month={Mar},
day={24},
volume={14},
number={1},
pages={1647},
issn={2041-1723},
doi={10.1038/s41467-023-37153-0},
}

@article{Paredes2024,
  author  = {Paredes, Rosario and Bruun, Georg M. and Camacho-Guardian, Arturo},
  title   = {Interactions mediated by atoms, photons, electrons, and excitons},
  journal = {Physical Review A},
  volume  = {110},
  pages   = {030101},
  year    = {2024},
  doi     = {10.1103/PhysRevA.110.030101}
}

@article{Ando2025Mediated,
  author  = {Ando, Henry and Cai, Geyue and Chin, Cheng and Enss, Tilman},
  title   = {Mediated interactions in mixtures of ultracold atoms},
  journal = {arXiv preprint arXiv:2511.09745},
  year    = {2025},
  doi     = {10.48550/arXiv.2511.09745}
}

@article{pes21,
  title = {Finite-range effects in the unitary Fermi polaron},
  author = {Pessoa, Renato and Vitiello, S. A. and Pe\~na Ardila, L. A.},
  journal = {Phys. Rev. A},
  volume = {104},
  issue = {4},
  pages = {043313},
  numpages = {8},
  year = {2021},
  month = {Oct},
  publisher = {American Physical Society},
  doi = {10.1103/PhysRevA.104.043313},
  url = {https://link.aps.org/doi/10.1103/PhysRevA.104.043313}
}

@Article{sch17,
  author    = {L. M. Schonenberg and G. J. Conduit},
  title     = {Effective-range dependence of resonant {F}ermi gases},
  journal   = {Physical Review A},
  year      = {2017},
  volume    = {95},
  number    = {1},
  month     = {Jan},
  doi       = {10.1103/physreva.95.013633},
  publisher = {American Physical Society ({APS})},
}

@Article{vli13,
  author       = {Jonas Vlietinck and Jan Ryckebusch and Kris Van Houcke},
  date         = {2013-03},
  year         = {2013},
  journal      = {Physical Review B},
  title        = {Quasiparticle properties of an impurity in a Fermi gas},
  doi          = {10.1103/physrevb.87.115133},
  number       = {11},
  pages        = {115133},
  volume       = {87},
  publisher    = {American Physical Society ({APS})},
}

@article{Perez2021,
author = {J. Pérez-Ríos},
title = {Cold chemistry: a few-body perspective on impurity physics of a single ion in an ultracold bath},
journal = {Mol. Phys.},
volume = {119},
number = {8},
pages = {e1881637},
year = {2021},
publisher = {Taylor \& Francis},
doi = {10.1080/00268976.2021.1881637},
}

@article{mysliwy2023longrange,
  title = {Long-range interacting Fermi polaron},
  author = {My\ifmmode \acute{s}\else \'{s}\fi{}liwy, Krzysztof and Jachymski, Krzysztof},
  journal = {Phys. Rev. B},
  volume = {109},
  issue = {21},
  pages = {214208},
  numpages = {9},
  year = {2024},
  month = {Jun},
  publisher = {American Physical Society},
  doi = {10.1103/PhysRevB.109.214208},
}

@article{Anderson1972MoreIsDifferent,
  author  = {P. W. Anderson},
  title   = {More is Different},
  journal = {Science},
  volume  = {177},
  number  = {4047},
  pages   = {393--396},
  year    = {1972},
  doi     = {10.1126/science.177.4047.393},
  url     = {https://doi.org/10.1126/science.177.4047.393}
}

@article{Cooper1956,
  author  = {Leon N. Cooper},
  title   = {Bound Electron Pairs in a Degenerate Fermi Gas},
  journal = {Physical Review},
  volume  = {104},
  number  = {4},
  pages   = {1189--1190},
  year    = {1956},
  doi     = {10.1103/PhysRev.104.1189},
  url     = {https://doi.org/10.1103/PhysRev.104.1189}
}

@article{BCS1957,
  author  = {J. Bardeen and L. N. Cooper and J. R. Schrieffer},
  title   = {Theory of Superconductivity},
  journal = {Physical Review},
  volume  = {108},
  number  = {5},
  pages   = {1175--1204},
  year    = {1957},
  doi     = {10.1103/PhysRev.108.1175},
  url     = {https://doi.org/10.1103/PhysRev.108.1175}
}

@article{Landau1957FermiLiquid,
  author  = {Lev D. Landau},
  title   = {The Theory of a Fermi Liquid},
  journal = {Soviet Physics JETP},
  volume  = {3},
  pages   = {920--925},
  year    = {1957},
  url     = {https://jetp.ras.ru/cgi-bin/dn/e_003_06_0920.pdf}
}

@article{Landau1957Oscillations,
  author  = {Lev D. Landau},
  title   = {Oscillations in a Fermi Liquid},
  journal = {Soviet Physics JETP},
  volume  = {5},
  pages   = {101--108},
  year    = {1957},
  url     = {https://jetp.ras.ru/cgi-bin/dn/e_005_01_0101.pdf}
}

@book{Nozieres1964FermiSystems,
  author    = {Pierre Nozi{\`e}res},
  title     = {Theory of Interacting Fermi Systems},
  publisher = {Benjamin},
  year      = {1964},
  url       = {https://www.worldscientific.com/worldscibooks/10.1142/0406}
}

@book{PinesNozieres1966QuantumLiquids,
  author    = {David Pines and Pierre Nozi{\`e}res},
  title     = {The Theory of Quantum Liquids},
  publisher = {W. A. Benjamin},
  year      = {1966},
  url       = {https://press.uchicago.edu/ucp/books/book/chicago/T/bo3633461.html}
}

@article{Schirotzek2009,
  author  = {A. Schirotzek and C.-H. Wu and A. Sommer and M. W. Zwierlein},
  title   = {Observation of Fermi Polarons in a Tunable Fermi Liquid of Ultracold Atoms},
  journal = {Phys. Rev. Lett.},
  volume  = {102},
  pages   = {230402},
  year    = {2009},
  doi     = {10.1103/PhysRevLett.102.230402},
  url     = {https://doi.org/10.1103/PhysRevLett.102.230402}
}

@article{Kohstall2012,
  author  = {C. Kohstall and M. Zaccanti and M. Jag and A. Trenkwalder and P. Massignan and G. M. Bruun and F. Schreck and R. Grimm},
  title   = {Metastability and coherence of repulsive polarons in a strongly interacting Fermi mixture},
  journal = {Nature},
  volume  = {485},
  pages   = {615--618},
  year    = {2012},
  doi     = {10.1038/nature11065},
  url     = {https://doi.org/10.1038/nature11065}
}

@article{Koschorreck2012,
  author  = {M. Koschorreck and D. Pertot and E. Vogt and B. Fr{\"o}hlich and M. Feld and M. K{\"o}hl},
  title   = {Attractive and repulsive Fermi polarons in two dimensions},
  journal = {Nature},
  volume  = {485},
  pages   = {619--622},
  year    = {2012},
  doi     = {10.1038/nature11151},
  url     = {https://doi.org/10.1038/nature11151}
}

@article{Nascimbene2009,
  author  = {S. Nascimb{\`e}ne and F. Chevy and C. Salomon and Y. Castin},
  title   = {Collective oscillations of an imbalanced Fermi gas: axial compression modes and polaron effective mass},
  journal = {Phys. Rev. Lett.},
  volume  = {103},
  pages   = {170402},
  year    = {2009},
  doi     = {10.1103/PhysRevLett.103.170402},
  url     = {https://doi.org/10.1103/PhysRevLett.103.170402}
}

@article{Massignan2014,
  author  = {P. Massignan and M. Zaccanti and G. M. Bruun},
  title   = {Polarons, dressed molecules and itinerant ferromagnetism in ultracold Fermi gases},
  journal = {Rep. Prog. Phys.},
  volume  = {77},
  pages   = {034401},
  year    = {2014},
  doi     = {10.1088/0034-4885/77/3/034401},
  url     = {https://doi.org/10.1088/0034-4885/77/3/034401}
}

@article{Prokofev2008,
  author  = {N. Prokof'ev and B. Svistunov},
  title   = {Fermi-polaron problem: Diagrammatic Monte Carlo method for divergent sign-alternating series},
  journal = {Phys. Rev. B},
  volume  = {77},
  pages   = {020408(R)},
  year    = {2008},
  doi     = {10.1103/PhysRevB.77.020408},
  url     = {https://doi.org/10.1103/PhysRevB.77.020408}
}

@article{Mora2009,
  author  = {C. Mora and F. Chevy},
  title   = {Ground state of the Fermi polaron},
  journal = {Phys. Rev. A},
  volume  = {80},
  pages   = {033607},
  year    = {2009},
  doi     = {10.1103/PhysRevA.80.033607},
  url     = {https://doi.org/10.1103/PhysRevA.80.033607}
}

@article{Combescot2009,
  author  = {R. Combescot and S. Giraud and X. Leyronas},
  title   = {Analytical theory of the dressed bound state in highly polarized Fermi gases},
  journal = {Europhys. Lett.},
  volume  = {88},
  pages   = {60007},
  year    = {2009},
  doi     = {10.1209/0295-5075/88/60007},
  url     = {https://doi.org/10.1209/0295-5075/88/60007}
}

@article{Sidler2017,
  author  = {M. Sidler and P. Back and O. Cotlet and A. Srivastava and T. Fink and M. Kroner and E. Demler and A. Imamoglu},
  title   = {Fermi polaron-polaritons in charge-tunable atomically thin semiconductors},
  journal = {Nat. Phys.},
  volume  = {13},
  pages   = {255--261},
  year    = {2017},
  doi     = {10.1038/nphys3949},
  url     = {https://doi.org/10.1038/nphys3949}
}

@article{Jastrow1955,
  author  = {Jastrow, Robert},
  title   = {Many-Body Problem with Strong Forces},
  journal = {Phys. Rev.},
  volume  = {98},
  pages   = {1479--1484},
  year    = {1955},
  doi     = {10.1103/PhysRev.98.1479}
}

@article{Foulkes2001,
  author  = {Foulkes, W. M. C. and Mitas, L. and Needs, R. J. and Rajagopal, G.},
  title   = {Quantum Monte Carlo simulations of solids},
  journal = {Rev. Mod. Phys.},
  volume  = {73},
  pages   = {33--83},
  year    = {2001},
  doi     = {10.1103/RevModPhys.73.33}
}

@article{Levinsen2025MediumEnhanced,
  title     = {Medium-Enhanced Polaron Repulsion in a Dilute Bose Mixture},
  author    = {Levinsen, Jesper and Bleu, Olivier and Parish, Meera M.},
  journal   = {Physical Review Letters},
  volume    = {135},
  number    = {11},
  pages     = {113402},
  year      = {2025},
  doi       = {10.1103/x2xb-6hzv},
  publisher = {American Physical Society}
}

@article{Ardila2022UltraDilute,
  title     = {Ultra-Dilute Gas of Polarons in a Bose--Einstein Condensate},
  author    = {Pe{\~n}a Ardila, Luis A.},
  journal   = {Atoms},
  volume    = {10},
  number    = {1},
  pages     = {29},
  year      = {2022},
  doi       = {10.3390/atoms10010029},
  publisher = {MDPI}
}

@article{Grusdt2025Impurities,
  title   = {Impurities and polarons in bosonic quantum gases: a review on recent progress},
  author  = {Grusdt, Fabian and Mostaan, Nils and Demler, Eugene and Ardila, Luis A. Pe{\~n}a},
  journal = {Reports on Progress in Physics},
  volume  = {88},
  number  = {6},
  pages   = {066401},
  year    = {2025},
  doi     = {10.1088/1361-6633/adc8a0},
  publisher = {IOP Publishing}
}

@article{Massignan2026Polarons,
  title   = {Polarons in atomic gases and two-dimensional semiconductors},
  author  = {Massignan, Pietro and Schmidt, Richard and Astrakharchik, Grigori E. and Imamoglu, Atac and Zwierlein, Martin and Arlt, Jan J. and Bruun, Georg M.},
  journal = {Reviews of Modern Physics},
  volume  = {98},
  number  = {3},
  pages   = {035002},
  year    = {2026},
  doi     = {10.1103/4nng-bb9z},
  publisher = {American Physical Society}
}

@article{Astrakharchik2004,
  author  = {Astrakharchik, G. E. and Boronat, J. and Casulleras, J. and Giorgini, S.},
  title   = {Equation of State of a Fermi Gas in the BEC-BCS Crossover: A Quantum Monte Carlo Study},
  journal = {Phys. Rev. Lett.},
  volume  = {93},
  pages   = {200404},
  year    = {2004},
  doi     = {10.1103/PhysRevLett.93.200404}
}

@article{RudermanKittel1954,
  author  = {Ruderman, M. A. and Kittel, C.},
  title   = {Indirect Exchange Coupling of Nuclear Magnetic Moments by Conduction Electrons},
  journal = {Physical Review},
  volume  = {96},
  pages   = {99--102},
  year    = {1954},
  doi     = {10.1103/PhysRev.96.99}
}

@article{Vivanco2025,
  author  = {F. J. Vivanco and A. Schuckert and S. Huang and G. L. Schumacher and G. G. T. Assump{\c{c}}{\~a}o and Y. Ji and J. Chen and M. Knap and N. Navon},
  title   = {The Strongly Driven Fermi Polaron},
  journal = {Nature Physics},
  volume  = {21},
  pages   = {564--569},
  year    = {2025},
  doi     = {10.1038/s41567-025-02799-8},
  url     = {https://doi.org/10.1038/s41567-025-02799-8}
}


\clearpage

\clearpage
\twocolumngrid

\vspace{1em}

\setcounter{section}{0}
\renewcommand{\thesection}{S\arabic{section}}

\setcounter{equation}{0}
\renewcommand{\theequation}{S\arabic{equation}}

\setcounter{figure}{0}
\renewcommand{\thefigure}{S\arabic{figure}}

\setcounter{table}{0}
\renewcommand{\thetable}{S\arabic{table}}

\section{End of Matter}

\subsection{I. Quantum Monte Carlo details -- Trial wave functions.}

The trial wave function satisfies the antisymmetry requirement of fermionic statistics and is constructed in Jastrow–Slater form~\cite{Jastrow1955,Foulkes2001,Astrakharchik2004}, which can be written as
\begin{equation}
    \Psi_T(R) = \prod_{i,i'}f(r_{i,i'}) D_{\downarrow}(N_{\downarrow}) D_{\uparrow}(N_\uparrow)
\end{equation}
where $f$ is the Jastrow factor that correlates unlike-spin pairs, and $i$ and $i'$ label the up- and down-spin particles, respectively. The nodal surface is completely defined by the product of the Slater determinants. The orbitals in the determinants are plane waves describing a Fermi sea of $N_\uparrow$ up spin particles and $N_\downarrow$ down spin impurities,
\begin{equation}
    D_{\downarrow}(N_{\downarrow}) D_{\uparrow}(N_\uparrow) = \left[ \mathcal{A} \prod_{j'}  e^{i \mathbf{k}\cdot \mathbf{r}_{j'}} \right ]\left[ \mathcal{A} \prod_{j}  e^{i \mathbf{k}\cdot \mathbf{r}_{j}} \right ] ,
\end{equation}
where $\mathcal{A}$ is an antisymmetrizer operator. For atoms confined in a cubic box of volume $\Omega = L^3$, the single-particle wave vector has modulus $k = \frac{2\pi}{L}\sqrt{n_x^2 + n_y^2 + n_z^2}$,where $n_{x},n_{y}$ and $n_{z}$ are integers. Typically, we consider a bath of up spin particles forming a closed-shell system, so that $N_\uparrow$ is defined by taking the maximum values of $I = n_x^2 + n_y^2 + n_z^2$. Therefore, we consider $N_\uparrow = 33, 57, 81, 93$ and $123$ particles which correspond to values of $I=4,5,6,8$ and $9$, respectively. The many-body trial wave function includes a Jastrow correlation factor, which captures the short-range correlations induced by the interaction. \\ 

All simulations, including those for the $N+1$ problem, the energy as a function of the impurity concentration, and the heavy-impurity case, were extrapolated to the thermodynamic limit. Finite-size effects were assessed by varying the number of particles in the Fermi bath while keeping the bath density fixed.

\subsection{II. Interaction Potential Models.}

\textbf{P\"oschl-Teller:} To describe the short-range interspecies interaction between the fermionic impurities and the particles of the Fermi bath, we employ the Pöschl-Teller (PT) attractive potential~\cite{car03,car05,gez09,for11,for12,gan11,gan14,cha04}, defined as
\begin{equation}
  V_{\text{PT}}(r) = -v_0 \frac{\hbar^2 \mu^2}{2 m_r} \frac{1}{\cosh^2(\mu r)} 
\end{equation}
where $m_r = \frac{m_\uparrow m_\downarrow}{m_\uparrow + m_\downarrow}$ is the reduced mass of the impurity-bath pair (corresponding to a mass ratio $m_\downarrow / m_\uparrow = 6.8$), $\mu$ sets the inverse range of the potential, and the dimensionless depth parameter $v_0$ tunes the strength of the interaction. The effective range is given analytically by $r_e = 2/\mu$. In our QMC simulations, $\mu$ is chosen sufficiently large such that $k_F r_e \ll 1$ (typically $k_F r_e \approx 0.03 - 0.32$), ensuring that the system remains strictly deep inside the universal short-range regime. The $s$-wave scattering length $a$ is tuned by varying $v_0$. The unitary limit corresponds exactly to $v_0 = 1$, where the potential supports its first zero-energy two-body bound state in vacuum. 

Unlike hard-core or square-well potentials, the Pöschl-Teller potential is infinitely differentiable across all space, avoiding sharp boundaries and large drift-force gradients $\mathbf{F} = 2 \boldsymbol{\nabla}\Psi_T / \Psi_T$ during Quantum Monte Carlo (QMC) sampling. 

The Jastrow factor $f(r)$ is derived using the Lowest Order Constrained Variational (LOCV) method~\cite{cow02,cha04,cha11,gan14}. In this approach, $f(r)$ at short distances is obtained by solving the relative two-body radial Schrödinger equation in vacuum:
\begin{equation}
-\frac{\hbar^2}{2m_r} \left( f''(r) + \frac{2}{r} f'(r) \right) + V_{\text{PT}}(r) f(r) = \lambda f(r), \quad (r < \bar{R})
\end{equation}
where $\lambda$ acts as an effective pair-interaction energy eigenvalue.

To ensure that interparticle correlations decay smoothly at large distances $f(r)$ is constrained to match a constant asymptote at a variational healing distance $\bar{R} \le L/2$: $f(r= \bar{R}) = 1$ and  
$\left. \frac{df(r)}{dr} \right|_{r = \bar{R}} = 0, \quad (r \ge \bar{R})$. The boundary conditions uniquely fix the eigenvalue $\lambda$ for a given healing length $\bar{R}$. The cutoff $\bar{R}$ is treated as an explicit variational parameter optimized within VMC to minimize the total expectation energy $\langle H \rangle_{\text{VMC}}$. This LOCV parametrization efficiently captures the exact two-body short-range physics while preserving the fixed-node structure set by the Slater determinants.\\

\textbf{Square-well:} For the interspecies interaction we adopt the square-well (SW) potential, a simple short-range model that has been extensively used in previous Quantum Monte Carlo studies of ultracold mixtures~\cite{ast04,lob06,ard15,sch17,pes21}. The potential is defined as
\begin{eqnarray}
 V(r) &=& 
 \begin{cases}
           -V_0, & r < R_0,  \\
            0,   & r > R_0,
 \end{cases}
 \label{eq:V_0SW}
 \end{eqnarray}
where $R_0$ is the interaction range and $V_0$ the depth of the well. The range $R_0$ is chosen sufficiently small to mimic the short-range nature of the interspecies interaction in the experiment~\cite{bar24}. In the zero-range limit, the model is fully characterized by the $s$-wave scattering length $a$, which we tune to explore both the weak- and strong-coupling regimes. 

Solving the two-body problem for the potential model described in \eq{V_0SW}, the scattering length is determined by the transcendental equation~\cite{ard15,ast04}  $a = R_0 \left[1 - \frac{\tan(K_0R_0)}{K_0R_0}\right]$, where $K_0=\sqrt{2\mu V_0/\hbar^2}$ with $\mu = \frac{m_\uparrow m_\downarrow}{m_\uparrow + m_\downarrow}$ the reduced mass. The unitary limit occurs at $K_0R_0 = \pi/2$. We choose $nR_0^3=10^{-6}$ sufficiently small to reach the universal short-range regime. Consequently, these calculations are intended as a zero-range benchmark rather than a quantitative model of the experimental resonance range.

\textbf{Zero-range:} With the aim of obtaining ground-state properties strictly in the zero-range limit without relying on finite-range model potentials and numerical extrapolations ($R_0 \to 0$), we employ the zero-range model governed by the Wigner–Bethe–Peierls boundary condition method~\cite{gio08,pes19}. When an unlike-spin pair separation $r_{i,j'}$ approaches zero, the many-body wave function $\Psi(\mathbf{R})$ satisfies the universal two-body short-distance asymptotic,
\begin{eqnarray}
    \lim_{r_{i,j'} \to 0} \Psi(\mathbf{R}) \propto \left( \frac{1}{r_{i,j'}} - \frac{1}{a} \right).
\end{eqnarray}

To satisfy the Wigner–Bethe–Peierls boundary condition at short distances while smoothly approaching unity at long distances, the two-body Jastrow correlation factor $f(r)$ for unlike-spin pairs is parametrized as~\cite{pes19}
\begin{equation}
f(r) = A \left[ \cosh(\lambda r) - \frac{r}{a} \right] \frac{1}{r}, \quad (0 < r < D) 
\end{equation}
where $D \le L/2$ is a variational healing length optimized in VMC to minimize the total energy, $A$ is a normalization constant satisfying $f(D) = 1$, and $\lambda$ is uniquely determined by enforcing the smooth matching condition at $r = D$: $\left. \frac{df(r)}{dr} \right|_{r=D} = 0$. For $r > D$, we set $f(r) = 1$, ensuring that particles do not interact with their periodic images across the simulation box boundaries.

\subsection{III. Simulation details}

We perform Variational Monte Carlo (VMC) and Fixed-Node Diffusion Monte Carlo (FN-DMC) calculations to determine the energy as a function of the impurity concentration $\mathcal{C} = N_\downarrow / N_\uparrow$. This concentration dependence allows us to extract both the single-polaron energy and the induced polaron-polaron interaction in the weak-to-moderate coupling regime where induced interactions scale linearly with $\mathcal{C}$. Trial wave function parameters are optimized within VMC before propagating in imaginary time using FN-DMC to project out the ground state within the fixed-node approximation.

To strictly control systematic errors in the FN-DMC calculations, key technical parameters were systematically tested and optimized. Production runs employ a typical imaginary-time step of $\Delta\tau = 1.0 \times 10^{-4} E_{F}^{-1}$. Convergence tests confirm that time-step discretization bias remains well within statistical uncertainties, matching the linear $\Delta\tau \to 0$ extrapolation limit. The configuration space is sampled using an average ensemble of $4000$ walkers, with population-control bias kept negligible by dynamically updating the reference energy $E_{\text{ref}}$. Furthermore, an initial imaginary-time equilibration period of at least $\tau E_F /\hbar \ge 7.0 $ is enforced to eliminate transient effects from trial wave function initialization before long-trajectory block averaging begins.

\subsection{IV. Local Inhomogeneity}

\begin{figure}[H] 
    \centering
    \includegraphics[width=\columnwidth]{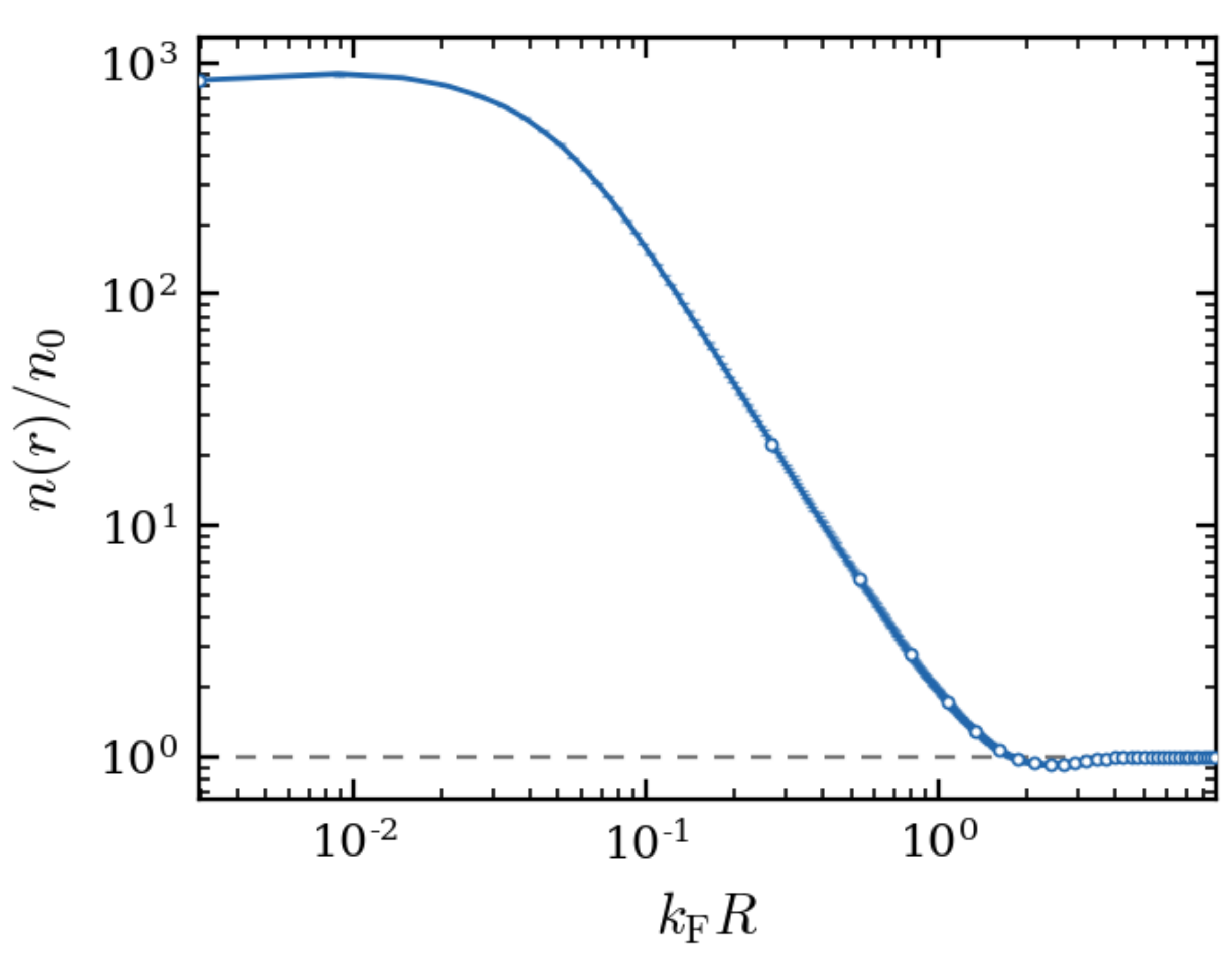}
    \caption{Normalized density in the strongly interacting regime, $X=0$ compute from the pair correlation function $g_{\uparrow\downarrow}(\mathbf{R})$.}
\label{Fig:FigS1}
\end{figure}

Quantum Monte Carlo methods provide direct access to the pair correlation function $g_{\uparrow\downarrow}(R)$ for mobile impurities, from which we can extract the density
profile $n(R)$ around the impurity as a function of the distance $R$, as shown in Fig.~\ref{Fig:FigS1}. We find that for distances $R \lesssim k_{\mathrm{F}}^{-1}$, the density is strongly inhomogeneous, leading to a substantial spatial variation of the local gas parameter in the strongly interacting regime. This pronounced short-range density inhomogeneity highlights an important advantage of the QMC approach, namely the strongly nonlinear rearrangement of the Fermi sea around the impurity is treated explicitly,
without relying on linear-response theory or local homogeneous-density approximations.

\end{document}